\documentclass[runningheads]{llncs}

\usepackage[T1]{fontenc}
\usepackage{amsmath}
\usepackage{amsmath}
\usepackage{bm}
\usepackage{amssymb}
\usepackage{caption}
\usepackage{booktabs}
\usepackage{microtype}
\usepackage[skip=4pt]{caption}
\usepackage[misc]{ifsym}
\usepackage{graphicx}
\usepackage{color}
\usepackage{multirow}
\begin{document}
\title{Controllable Surface Diffusion Generative Model for Neurodevelopmental Trajectories}
\titlerunning{Controllable Surface Diffusion Model}%
%
\author{Zhenshan Xie \and Levente Baljer \and M. Jorge Cardoso \and Emma Robinson\textsuperscript{(\Letter)}}
\authorrunning{Z. Xie et al.}
\institute{
Research Department of Biomedical Computing,\\ School of Biomedical Engineering \& Imaging Sciences,
King's College London\\
\email{\{emma.robinson\}@kcl.ac.uk}
}
%
%
%
\maketitle              
\begin{abstract}
Preterm birth disrupts the typical trajectory of cortical neurodevelopment, increasing the risk of cognitive and behavioral difficulties. However, outcomes vary widely, posing a significant challenge for early prediction. To address this, individualized simulation offers a promising solution by modeling subject-specific neurodevelopmental trajectories, enabling the identification of subtle deviations from normative patterns that might act as biomarkers of risk. While generative models have shown potential for simulating neurodevelopment, prior approaches often struggle to preserve subject-specific cortical folding patterns or to reproduce region-specific morphological variations. In this paper, we present a novel graph-diffusion network that supports controllable simulation of cortical maturation. Using cortical surface data from the developing Human Connectome Project (dHCP), we demonstrate that the model maintains subject-specific cortical morphology while modeling cortical maturation sufficiently well to fool an independently trained age regression network, achieving a prediction accuracy of $0.85 \pm 0.62$. 
\end{abstract}
\section{Introduction}
The human cerebral cortex is a highly convoluted sheet of grey matter that is best modelled as a surface~\cite{robinson2014msm,fischl1999high,glasser2016multi}. As one of the most highly evolved areas of the brain relative to non-human primates, cortical function underpins many aspects of higher-order cognition and is implicated in many neurological and psychiatric disorders \cite{paus2008many,roe2021asymmetric}. Unfortunately, localising signs of pathology in individual brains can be extremely challenging due to the scale of variation of human cortical organisation ~\cite{glasser2016multi,kong2019spatial,gordon2017precision}, which obscures subtle signs of disease. As such, methods that can accurately disentangle healthy variation from pathology in cortical anatomy would enable earlier and more precise detection of at-risk individuals.

One popular approach that has gained recent prominence is normative modeling. The objective here is to learn a generative model that encodes how brains vary normally relative to continuous phenotypes, such as age or behavioural scores. Classical normative models of cortical variation have been fit with Gaussian Process Regression (GPR) \cite{marquand2016understanding,rutherford2022normative} or Generative Additive Models of Location Scale and Shape (GAMLSS) ~\cite{bethlehem2022brain}. However, these rely on traditional image processing pipelines that combine diffeomorphic image registration with propagation of cortical features from individuals to population average templates~\cite{yang2020sample}. While diffeomorphic registration ensures smooth mappings, it underestimates the scale of human macro- and micro-structural cortical heterogeneity, leaving residual sources of variability that are addressed through smoothing. This process tends to mask subtle, subject-specific signatures of disease and cognition, especially in regions with high anatomical variability. 

Deep generative models offer a powerful alternative, as they can be trained to be invariant to residual misalignments of the data. Various architectures have been employed for this purpose, including VAEs~\cite{bass2022icam,ravi2022degenerative}, flow-based models~\cite{hwang2019conditional}, and Generative Adversarial Networks (GANs)~\cite{bai2022novel}. In this setting, the task can be reframed as an image-to-image translation problem, where the objective of the generator is to alter the appearance of an input image to change its neurological phenotype, such as regional brain volumes, or tissue intensity characteristics associated with developmental stages or disease states. Direct synthesis of translated images, however, risks changing the underlying brain anatomy of each individual.  Recent studies have therefore instead focused on learning deformation fields or difference maps ~\cite{bass2022icam}, constrained to translate images in ways that encode biologically-plausible, temporal or pathological changes, while preserving individual anatomical traits. Although previous research achieved good results on volumetric data, few attempts have been made to adapt these methods to non-Euclidean domains such as cortical surfaces~\cite{fawaz2021benchmarking}. \cite{fawaz2023surface} proposed a GAN-based network for simulating cortical neurodevelopment, demonstrating potential for the preservation of surface morphologies; however, this method is subject to unstable training dynamics due to adversarial learning. To this end, denoising Diffusion Models~\cite{ho2020denoising} have recently emerged as powerful generators of natural images with the ability to produce high-fidelity outputs through a stable training process. Although recent work~\cite{liu2023meshdiffusion,huang2022riemannian} has successfully adapted this framework to simple manifolds and low-resolution meshes, such methods are not suitable for modeling high-resolution cortical surface grids.

To address this problem, we propose a novel graph-diffusion network for the conditional generation of cortical neurodevelopment, capable of modelling postmenstrual age at scan (PMA), using data from the developing Human Connectome Project (dHCP). To enable the simulation of subject-specific patterns of cortical maturation, we adapt ControlNet~\cite{zhang2023adding} - leveraging its capabilities for external conditioning of inputs - to flexibly control a pretrained graph-diffusion model. We show that this approach preserves individual cortical morphologies to generate biologically-plausible trajectories of cortical maturation. This ability of our model to tailor predictions of typical neurodevelopment to the brains of individual patients offers potential for improved risk stratification and early diagnosis of neurodevelopmental conditions. 

\section{Methods}
\begin{figure}
    \centering
    \includegraphics[width=1.0\linewidth]{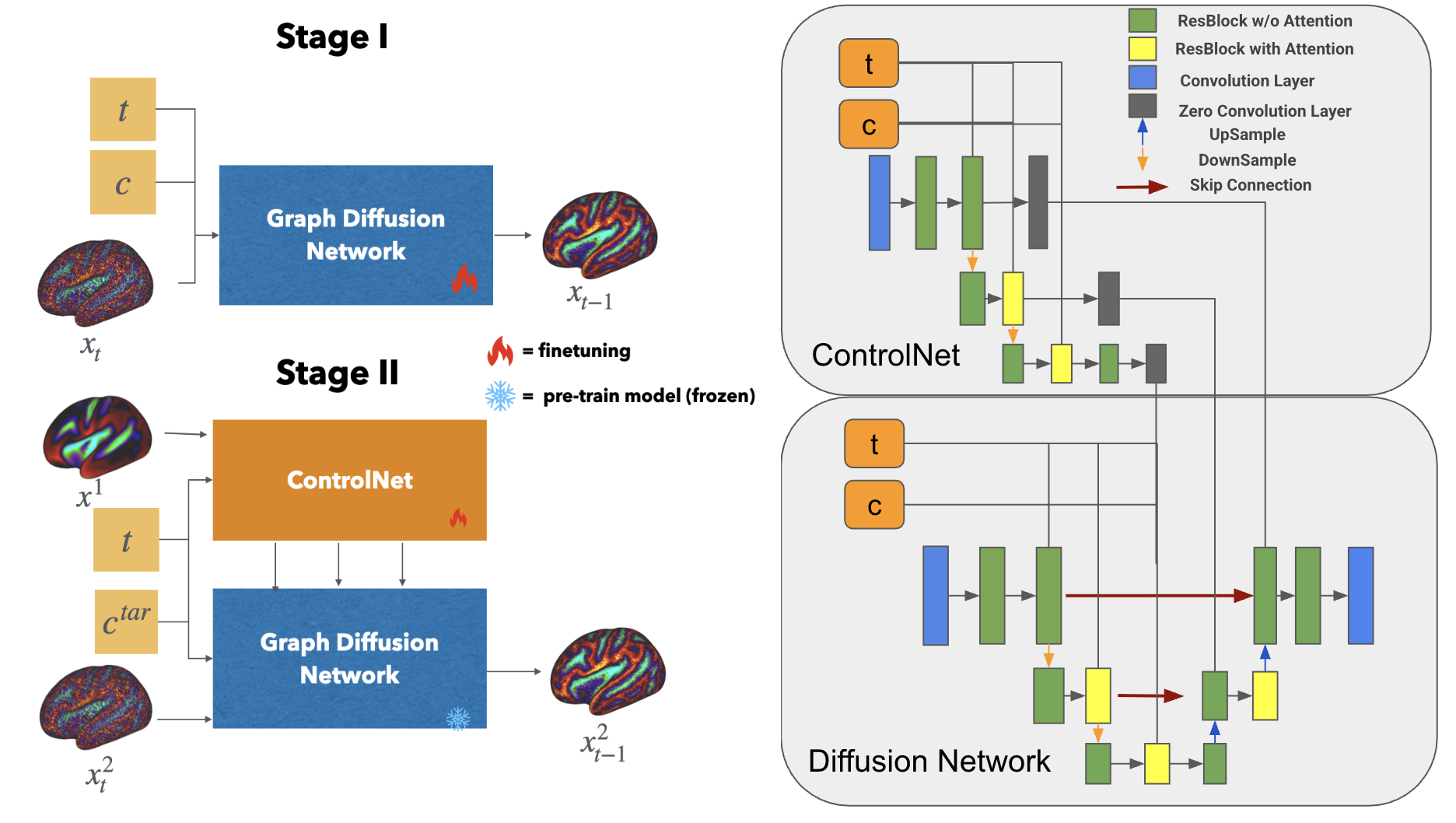}
    \caption{The figure shows the framework of the two-stage controllable cortical surface diffusion network. In stage I, we pre-train a Graph Diffusion Network to simulate cortical maturation, conditioned on time step $t$ and additional control conditions. In stage II, a ControlNet architecture guides the generation process while keeping the pre-trained Graph Diffusion Network parameters frozen.}
    \label{fig:Model}
\end{figure}

The inputs to our model are sphericalised, icosahedral cortical-feature maps, representing sulcal depth at resolution 40962 vertices. The objective of the model is to continuously simulate phenotypic changes in sulcal depth, whilst preserving overall anatomy. This is achieved through graph-based diffusion with the use of a ControlNet for tailored, subject-specific conditioning.  All components are summarized in Fig.~\ref{fig:Model}

\subsection{Background - Diffusion Models}
DDPM~\cite{ho2020denoising} models the data distribution $p(x)$ through a two-stage generative modelling process. This starts with forward diffusion – implemented as a Markov chain that progressively corrupts the original data $x_0 \sim p(x)$ with Gaussian noise, added over T steps:
\begin{equation}
\label{eq:add_noise}
    q\left(x_t \mid x_{t-1}\right)  =\mathcal{N}\left(x_t ; \sqrt{1-\beta_t} x_{t-1}, \beta_t \boldsymbol{I}\right)
\end{equation}
Here $\beta_t \in(0,1)$ scales the mean and variance of the distribution. 
Denoising is then framed as a learning problem, where a neural network $\epsilon_\theta$ is trained to iteratively predict the noise added at each diffusion step. 
In practice, this is optimised through minimising a mean-squared error (MSE) between the predicted $\epsilon_\theta\left(x_t\right)$ and sampled ($\epsilon_t$) Gaussian noise:
\begin{equation}
\label{eq:loss}
L={ }_{t \sim[1, T], x_0, \epsilon_t}\left[\left\|\epsilon_t-\epsilon_\theta\left(\sqrt{\bar{\alpha}_t} x_0+\sqrt{1-\bar{\alpha}_t} \epsilon_t, t\right)\right\|\right]
\end{equation}
Here $\bar{\alpha}_t=\prod_{i=1}^t\left(1-\beta_i\right)$. Once trained, the model can generate new samples from pure noise $x_T \sim \mathcal{N}(0, \boldsymbol{I})$ by iteratively applying the learned denoising process. In 2D and 3D diffusion is implemented by passing images through convolutional U-Nets, conditioned on diffusion iteration time ($t$). 

\subsection{Graph-Based Diffusion Model - Stage I}
To adapt the diffusion process to the spherical domain, we employ the graph convolution method from~\cite{fawaz2023surface}. This approach operates on spherical meshes with varying resolutions (e.g., 40,962, 10,242, or 2,562 vertices) and defines convolution as a weighted sum over each vertex's 1-ring neighborhood~\cite{morris2019weisfeiler}:
\begin{equation}
\hat{x}^i = W_1 x^i + W_2 \sum_{j \in N(i)} e_{i,j} x^j
\end{equation}
Here $x^i$ represents the feature vector of vertex $i$, $\hat{x}^i$ is the updated feature after convolution, $e_{i,j}$ denotes the edge weight from the adjacency matrix, $W_1$ and $W_2$ are learned filter weights.

Building upon this graph convolution foundation, we implement a U-Net architecture specifically designed for spherical data. The network consists of multiple levels, each containing 2 ResBlocks, except the bottleneck level, which uses 3. Each ResBlock incorporates the diffusion timestep ($t$) through sinusoidal positional encoding followed by an MLP, as described in~\cite{ho2020denoising}. Downsampling is performed through hexagonal pooling \cite{zhao2019spherical}, while upsampling is implemented using bilinear interpolation \cite{fawaz2023surface}.

To enable phenotype-conditioned generation, we integrate cross-attention modules into the final two network levels. The cross-attention mechanism follows:
\begin{equation}
\begin{gathered}
\text { Attention }(x, c)=\operatorname{softmax}\left(\frac{Q K^{\top}}{\sqrt{d_k}}\right) V  \\  Q=x W^Q, \quad K=c W^K, \quad V=c W^V
\end{gathered}
\end{equation}
where x represents the spatial feature map, where the attention is applied across the channel-wise embeddings at each spatial location, $c$ denotes the conditioning information, and $W^Q$, $W^K$, $W^V$ are learned linear projection matrices.

In the first stage, we train the Graph Diffusion Network to model the generative process of cortical surface conditioned on covariates $c = \langle s\rangle$, where $s$ is the postmenstrual scan age, following the loss defined by Eq.\eqref{eq:loss}. Once trained, these parameters are frozen for Stage II ControlNet training.


\subsection{ControlNet - Stage II}
Although effective in learning population-level developmental trends, attention-based conditioning is insufficient to preserve individual anatomical characteristics. This limitation arises because attention mechanisms primarily modulate feature importance at a coarse level without directly enforcing spatial alignment or vertex-wise consistency. As a result, fine-grained cortical morphologies tend to drift during the diffusion process, a phenomenon also observed in natural image generation tasks. 
ControlNet~\cite{zhang2023adding} addresses the issue by incorporating an additional network that encodes the key structural properties of inputs, and inserts this into the bottleneck of the diffusion module to guide generation of outputs. We conceptualize ControlNet and the graph diffusion model as a unified network $\epsilon_{\theta,\phi}$, where $\theta$ represents the frozen parameters of the diffusion network and $\phi$ denotes the trainable network's parameters of ControlNet, with $\phi$ initialized from $\theta$ to ensure compatibility and consistent feature extraction between both networks.

\subsubsection{Training} ControlNet leverages pairs of longitudinal cortical surfaces $(x^{1}, x^{2})$, acquired from the same individual at different ages. The network takes the $x^{1}$ as input, conditioned on the target phenotype $c^{2}$, and learns to match the noise residual (predicted by the diffusion model) of the follow-up surface $x^{2}$. The overall training objective seeks to minimize an MSE loss:
\begin{equation}
    \mathcal{L}=\mathbb{E}_{ t,x^{1}, x^{2}, c^2, \epsilon \sim \mathcal{N}(0,1)}\left[\left\|\epsilon_t-\epsilon_{\theta,\phi}\left(x^{1}, x_t^{2}, c^{2}, t\right)\right\|_2^2\right]
\end{equation}
Here, $x_t^{2}$ is the noisy version of the target surface at diffusion step $t$, and $\epsilon_t$ is the ground-truth noise for $x^2$. 

\subsubsection{Inference}
At test time, the model can continuously change the appearance of cortex $x'$ to match any given target covariates: $c^{tar} = \langle s^{tar}\rangle$ by presenting these to the ControlNet. This guides generation by injecting subject specificity and conditioning into the denoising process, synthesising a target surface that not only conforms to the population-level phenotypic trends, but also preserves individual cortical morphologies encoded from $x'$. This inference strategy supports anatomically plausible and developmentally consistent simulations of cortical maturation tailored to each individual.

\section{Experiments}
\begin{figure}
    \centering
    \includegraphics[width=0.95\linewidth]{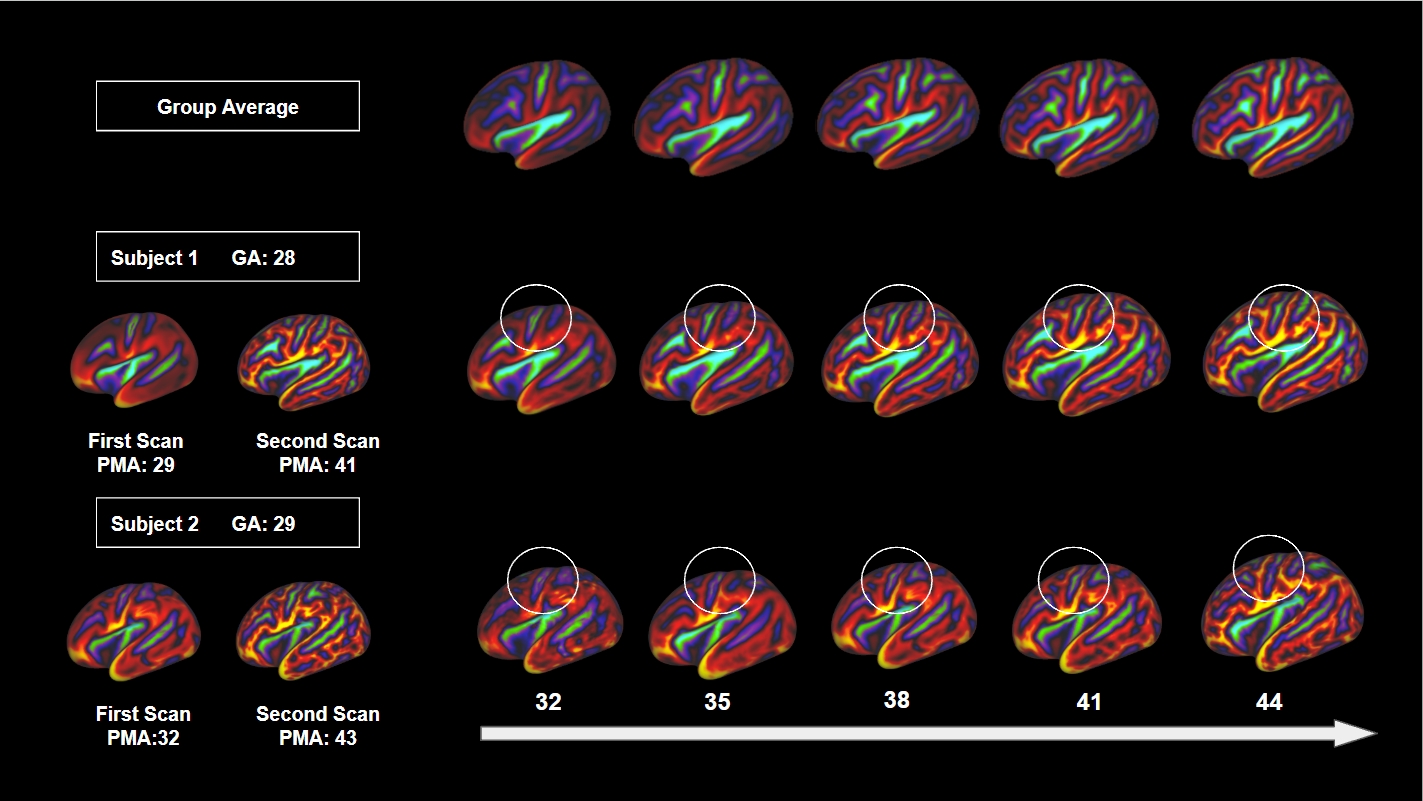}
    \caption{The figure shows the cortical maturation patterns across PMA. The top row shows the group-average cortical surfaces at five representative PMA(32,35,38,41,44 weeks). The left panel of the middle and bottom rows shows two individual subjects scanned twice and their corresponding GA and PMA. The right panel shows the cortical surfaces generated by the proposed diffusion model, simulating typical cortical maturation across the PMA. The white circle indicates the frontal lobe area.}
    \label{fig:enter-label}
\end{figure}
\subsection{Dataset}
All experiments used data from the developing Human Connectome Project (dHCP), 
comprising 681 subjects in total, cross-sectionally scanned between 24-45 weeks post-menstrual age (PMA), and gestational ages at birth (GA) between 21-40 weeks. In the dHCP study, the preterm group typically undergoes a scan either shortly after birth or at term-equivalent age; we also leverage a subset of 90 subjects scanned twice at both timepoints. Cortical surface meshes and sulcal depth metrics were derived from the dHCP structural pipeline described in~\cite{edwards2022developing,makropoulos2018developing}. Similar to FreeSurfer~\cite{fischl2012freesurfer}, this projects all data onto spheres; following which all maps were then registered to the dHCP 40-week neonatal sulcal depth template from the dHCP spatiotemporal cortical surface atlas~\cite{bozek2018construction} using multimodal surface matching (MSM)~\cite{robinson2014msm,robinson2018multimodal}, driven by sulcal depth features; then resampled to a regular 40,962-vertex icosphere using barycentric interpolation, implemented using Human Connectome Project (HCP) workbench~\cite{marcus2011informatics}. In this paper, we only used the sulcal depth feature maps as our input channels.

\begin{figure}
    \centering
    \includegraphics[width=0.5\linewidth]{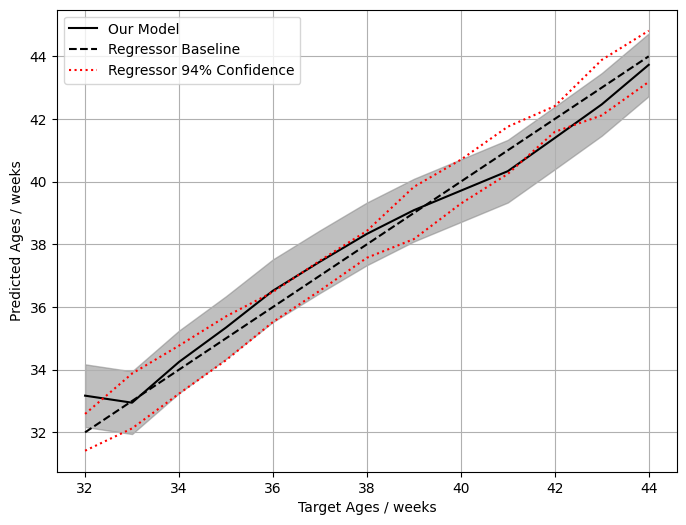}
    \caption{A comparison of age prediction accuracy for simulated and ground truth cortical feature maps. The horizontal line represents the target PMA, and the vertical line shows the PMA predicted by the independently trained SiT~\cite{dahan2022surface}. The shaded grey area reflects the uncertainty range of the SiT model’s predictions. The red dashed lines depict the $ 94.1\%$ confidence interval of the SiT.}
    \label{fig:line}
\end{figure}
\vspace{-5mm}

\subsection{Implementation Details}
Feature maps were augmented offline with non-linear warps and rotations to produce realistic variations in the data and improve network generalisation \cite{fawaz2023surface}; these were then normalized to range (-1,1) using symmetric normalization. For the training of stage I, we used the subjects without follow-up scans; the training and validation sets were allocated in the ratio of 473:59:59 examples (i.e. $\sim 80\%:10\%:10\%$). The diffusion network was trained with an AdamW Optimizer with a learning rate of 0.0001 for 1000 epochs. For the training of stage II, we used the subset of 90 subjects who were scanned twice. Due to the limited amount of data, we used 5-fold cross-validation, splitting the data into training and validation sets of 72 and 18 (i.e. 80\% and 20\%). The ControlNet was trained with an AdamW Optimizer with a learning rate of 0.00005 for 500 epochs.

\subsection{Evaluation}

\noindent \textbf{Age Regressor Evaluation:} Evaluation was framed from two perspectives: firstly, we quantified whether the model generated realistic images of the targeted age by performing age regression using an independently trained SiT~\cite{dahan2022surface}. The SiT regression model was trained with the same dataset as diffusion models, which can regress apparent age from the ground truth sulcal depth maps. This returned a baseline age MAE of $0.59 \pm 0.51$ weeks PMA. We then synthetically modified the age of all test examples - generating examples at each week from 32 to 44 weeks- and regressed their age from the trained model. 
\begin{table}[htbp]
\centering
\captionsetup{skip=2pt} 
\setlength{\tabcolsep}{16pt}
\begin{tabular}{|l|c|}
\hline
\textbf{Model} & \textbf{PMA Accuracy} \\
\hline
CycleGAN & $7.66 \pm 2.11$\\
\hline
3-cycle~\cite{fawaz2023surface} & $1.02 \pm 0.2$ \\
\hline
ours & $0.85 \pm 0.62$\\
\hline
\end{tabular}
\caption{A table comparing the performance of different generative models on the simulation of typical cortical developmental trajectories, as measured by the accuracy of synthesised cortical surfaces to target PMA and subject specificity.}
\label{tab:age_accuracy}
\end{table}
\vspace{-20pt}

\noindent \textbf{Follow-up Scan Comparisons:}
Secondly, we evaluated the extent to which the model preserves subject specificity by leveraging the longitudinal data and comparing the similarity of the simulated images with their ground truth follow-up, using three quantitative metrics: Mean Squared Error (MSE), Peak Signal-to-Noise Ratio (PSNR), and Structural Similarity Index Measure (SSIM).

\subsection{Results}
The results in Table~\ref{tab:age_accuracy} show that the data generated by our model achieved an age MAE of $0.85 \pm 0.62$, outperforming the baseline network~\cite{fawaz2023surface}. Compared to the baseline regression performance on ground truth data, our model demonstrates comparable accuracy, though slightly larger errors are observed at the extremes of the age distribution, likely due to limited data availability in those ranges. These findings suggest that the proposed model can effectively generate synthetic cortical surfaces that accurately reflect age-related morphological trends. 
To further validate identity preservation, we conducted statistical comparisons between each preterm individual’s follow-up scan and their simulated images. The results show that the model can maintain a high degree of structural similarity. In addition, Fig.~\ref{fig:enter-label} presents qualitative examples of the generated cortical surfaces across time for two representative preterm individuals. The white circle highlights the frontal lobe area of the cortical surface, which can be confirmed to contain subject-specific information. Comparisons with population-average templates indicate that the changes in sulcal depth align with expected neurodevelopmental patterns. As reported in Table~\ref{tab:follow-up}, our model achieves the lowest Mean Squared Error (MSE) of 0.037, significantly outperforming the baseline model. These results collectively suggest that our model not only provides accurate age-conditioned synthesis but also better preserves subject identity compared to existing methods.


\begin{table}[htbp]
\centering
\captionsetup{skip=2pt} 
\setlength{\tabcolsep}{16pt} 
\begin{tabular}{|l|c|c|c|}
\hline
\textbf{Model} & MSE & PSNR& SSIM \\
\hline
CycleGAN & 0.203 & 17.4 &0.73\\
\hline
3-cycle~\cite{fawaz2023surface} & 0.184 & 17.9 & 0.80\\
\hline
ours & 0.037 & 25.6 & 0.82 \\
\hline
\end{tabular}
\caption{The results for PMA-Conditioned follow-up scan comparison.}
\label{tab:follow-up}
\end{table}
\vspace{-35pt}

\section{Conclusion}
In this work, we present a novel deep graph-based diffusion generative model for simulating neonatal cortical surface neurodevelopment. Our approach addresses key challenges in surface-based modelling by integrating geometric deep learning with Euclidean diffusion frameworks, enabling the generation of anatomically precise, age-consistent cortical surfaces. Notably, the model preserves subject-specific cortical features while accurately capturing neurodevelopmental trajectories. This contributes to improving the individualization of prematurity neurodevelopmental simulations, which is essential for early risk stratification. Future directions include extending the model to incorporate fetal imaging to better capture early developmental stages and to mitigate the assumption that near-birth preterm scans fully represent healthy development. Additionally, conditioning on discrete demographic and clinical factors such as disease status, sex, and birth weight could further enhance the model's ability to disentangle the complex influences on early brain development. These advancements would strengthen the model’s potential as a valuable tool for personalized neurodevelopmental assessment and early intervention planning.

\vspace{1em}

\noindent\textbf{Acknowledgments.}The dHCP neonatal dataset was provided by the developing Human Connectome Project, KCL-ImperialOxford Consortium funded by the European Research Council (ERC) under the European Union Seventh Framework Programme (FP/2007-2013) / ERC Grant Agreement no. [319456]. ECR acknowledges philanthropic support from Heart of Racing LLC to the Brain Health in Gen2020 programme at King’s College London and support from the Medical Research Council (UKRI534). ZX acknowledges funding from King's-China Scholarship(K-CSC) PhD Scholarship program.
\vspace{1em}

\noindent\textbf{Disclosure of Interests.} The authors have no competing interests to declare that are relevant to the content of this article.
%
%
\bibliographystyle{splncs04}

%

\end{document}